\documentclass[12pt]{article}
\pdfoutput=1
\usepackage{amsmath,amssymb,bbold,epsfig}
\usepackage{amsfonts}
\usepackage{slashed}
\usepackage{dsfont}
\usepackage{cite}
\usepackage{graphicx,color}
\usepackage{bbm}
\usepackage[usenames,dvipsnames]{xcolor}
\definecolor{lightgray}{gray}{0.90}
\usepackage[colorlinks=true,linkcolor=red,citecolor=blue,urlcolor=cyan,
bookmarks=true,bookmarksopen=true,pdfpagemode=None,pdfstartview=FitH]{hyperref}

\newcommand{\be}{\begin{equation}}
\newcommand{\ee}{\end{equation}}
\newcommand{\bea}{\begin{eqnarray}}
\newcommand{\eea}{\end{eqnarray}}

\begin{document}

\title{
\vglue 1.3cm
\Large \bf
On chirality and chiral neutrino oscillations} 
\author{
\vspace*{-0.2cm}
{Evgeny~Akhmedov
\!\thanks{Email: \tt 
akhmedov@mpi-hd.mpg.de} 
\vspace*{5.5mm}
} \\
{\normalsize\em 
Max-Planck-Institut f\"ur Kernphysik, Saupfercheckweg 1, 
}\\
{\normalsize\em
69117 Heidelberg, Germany
\vspace*{0.85cm}} 
}
\date{}  

\maketitle 
\thispagestyle{empty} 
\vspace{-0.8cm} 
\begin{abstract} 
It has been claimed in a number of publications that neutrinos can 
exhibit chirality oscillations. In this note we discuss the notion of 
chirality and show that chiral neutrino oscillations in vacuum do not 
occur. We argue that the incorrect claims to the contrary resulted from 
a failure to clearly discriminate between quantum fields, states and 
wave functions. We also emphasize the role played in the erroneous 
claims on the possibility of chirality oscillations by the widely spread 
misconceptions about negative energies.
\end{abstract}
\vspace{0.2cm}
\vspace{0.3cm}

\newpage


\section{\label{sec:intro}Introduction} 
Charged-current weak interaction have chiral ($V-A$) structure, and therefore 
neutrinos produced in the corresponding processes,  such as 
e.g.\ nuclear $\beta$-decay, are often said to be chirality eigenstates: 
$\nu_L$ for neutrinos and $\bar{\nu}_R$ for antineutrinos if neutrinos are 
Dirac particles, or $\nu_L$ and their CPT-conjugates $\nu^c_R=(\nu_L)^c$ 
if they are of Majorana nature. As the chirality projectors $P_L=\frac{1}{2}
(1-\gamma_5)$ and $P_R=\frac{1}{2}(1+\gamma_5)$ do not commute with the Dirac 
Hamiltonian $H_D=\gamma^0\vec{\gamma}\vec{p}+\gamma^0 m$, it is often said 
that chirality is not a good quantum number for free fermions of nonzero 
mass, i.e.\ it is not conserved; hence, chiral states are not eigenstates of 
free propagation. This appears very similar to the situation with neutrinos 
of definite flavour: neutrino states produced in charged-current processes 
are flavour eigenstates, which do not coincide with the mass eigenstates, 
i.e.\  they are not eigenstates of free propagation. Instead, they are 
linear superpositions of neutrino states 
of definite mass. With time (or traveled distance) the different mass 
components of the originally produced flavour state develop different phases, 
and the phase differences lead to neutrino flavour oscillations.  
This analogy has prompted a number of authors to conclude that 
neutrino chirality can also oscillate (see ref.~\cite{Smirnov:2025wax} for an 
extensive list of literature). Neglecting the existence of more than one 
neutrino species and the related flavour issues (which are not directly 
relevant to the alleged chirality oscillations), for the probabilities of 
left-handed neutrinos to become right-handed or to stay left-handed upon the 
time $t$ after the neutrino production they found, respectively, 
\be
P(\nu_L\to\nu_R,t)=\frac{m^2}{E^2}\sin^2(Et)\,,\qquad\quad
P(\nu_L\to\nu_L,t)=1-\frac{m^2}{E^2}\sin^2(Et)\,,
\label{eq:wrong1}
\ee
where $E=E_p\equiv +\sqrt{\vec{p}\,^2+m^2}$. The derivations of these 
relations were based on the observation that, unlike the usual spinors $u$ 
and $v$, which 
are the solutions of the Dirac equation in momentum space and eigenspinors of 
the Dirac Hamiltonian $H_D$, their chiral projections, such as e.g.\ 
$u_L\equiv P_Lu$, are not. The expansion of $u_L$ in the full basis of the 
eigenspinors of $H_D$ unavoidably includes the spinors of $v$-type, which 
correspond to negative-energy solutions of the Dirac equation. The phase 
$(Et)$  of the oscillatory factors in eq.~({\ref{eq:wrong1}) then comes from 
the differences of the phases corresponding to the evolution of the 
positive-energy and negative-energy terms in the expansion of $u_L$: 
\,$\Delta\phi=\frac{1}{2}[E-(-E)]t$. 

It has been demonstrated in the recent paper~\cite{Smirnov:2025wax} that these 
results are incorrect, and the oscillations of neutrino chirality in vacuum do 
not take place. While we fully agree with this conclusion 
of~\cite{Smirnov:2025wax}, in this note we consider the issue of chiral 
oscillations from a somewhat different perspective, emphasizing its relations 
with some basic notions of quantum field theory (QFT). 

Before proceeding, we note that eq.~(\ref{eq:wrong1}) cannot actually 
describe the probabilities of any physical processes. The point is that the 
oscillation phase $Et$ is not Lorentz invariant, and therefore 
neither are the probabilities of chiral oscillations. 
However, in a Lorentz invariant theory the probabilities of all physical 
processes must be invariant with respect to Lorentz transformations. 
As an example, the standard expressions for the probabilities of flavour 
oscillations in vacuum are Lorentz invariant: they depend on 
the phases of the form $\frac{\Delta m^2}{2E}t$ (if neutrino 
evolution in time is considered) or $\frac{\Delta m^2}{2p}L$ (if 
one is interested in the evolution with the distance traveled by the 
neutrinos). These oscillation probabilities 
actually apply when neutrinos can be considered as 
pointlike objects \cite{Akhmedov:2009rb}, and 
it is easy to show that in this approximation the quantities $t/E$ and $L/p$ 
are Lorentz invariant \cite{Levy:2009uz,Akhmedov:2009rb}.%
\footnote{
It is often stated in the literature that the neutrino oscillation 
probabilities in vacuum depend on $L/E$. However, careful derivation 
actually yields the dependence on $L/p$ \cite{Akhmedov:2009rb}. 
While for relativistic neutrinos $L/E$ is essentially the same as $L/p$, when 
considering the Lorentz invariance issues it is important to distinguish 
between them, because  
$L/p$ is Lorentz invariant for pointlike particles,  
whereas $L/E$ is not.
}
The quantity $Et$ is {\sl not} Lorentz invariant, irrespective of whether 
neutrinos are assumed to be pointlike or not.

\section{\label{sec:QFT}QFT considerations} 

So, what could have gone wrong with the derivations of the expressions in 
eq.~(\ref{eq:wrong1})? We shall examine this issue within the formalism of 
QFT. Its basic elements are quantum fields, Lagrangians built out of these 
fields and states (state vectors)%
\footnote{In this note we use the terms ``state" and ``state vector" 
interchangeably. Where it cannot cause confusion, we also use the term 
neutrino for both neutrinos and antineutrinos.}. 
Wave functions can be calculated but do not play any significant role in QFT. 
As we shall see, it is the confusion of states, fields and wave functions, 
often caused by the use of the same notation for these very different 
objects, that led some authors to the incorrect conclusion about the 
possibility of chiral oscillations.%
\footnote{This confusion is unfortunately rather widespread and is not limited 
to the issue of chiral neutrino oscillations. However, to discuss it outside 
this context goes beyond the scope of the present note.}  

To set the stage, consider a $\beta^-$-decay process 
$A(Z,N)\to A(Z+1,N-1)+e^-+\bar{\nu}_e$. The leptonic charged current 
responsible for this process is 
\be
j^\mu(x)=\bar{e}(x)\gamma^\mu(1-\gamma_5)\nu(x)\,.
\label{eq:leptCC}
\ee
Here the electron quantum field operator can be expanded as 
\be
e(x)=\int\frac{d^3p}{(2\pi)^3\sqrt{2E_{e\vec{p}}}}\sum_s \left[
b_e(p,s) u_e(p,s) e^{-i p x} + d_e^\dag(p,s) v_e(p,s) 
e^{i p x}\right].\quad\quad~\;\,
\label{eq:decompE}
\ee
For Dirac neutrinos, the quantum field operator can be written in a similar 
form:  
\be
\nu(x)=\int\frac{d^3p}{(2\pi)^3\sqrt{2E_{\nu\vec{p}}}}\sum_{s}
\left[b_\nu(p,s)u_\nu(p,s) e^{-i p x} +
d^\dag_\nu(p,s) v_\nu(p,s) e^{i p x}\right],
\label{eq:decompNuD}
\ee
whereas for Majorana neutrinos, 
\be
\nu(x)=\int\frac{d^3p}{(2\pi)^3\sqrt{2E_{\nu\vec{p}}}}\sum_{s}
\left[a_\nu(p,s)u_\nu(p,s) e^{-i p x} +
a^\dag_\nu(p,s) v_\nu(p,s) e^{i p x}\right]. 
\label{eq:decompNuM}
\ee
For simplicity, we don't differentiate  
between neutrino mass and flavour states, as the leptonic mixing is 
extraneous to the issues we are discussing. It can, however, be readily 
incorporated.   
The spin sums in eqs.~(\ref{eq:decompE})-(\ref{eq:decompNuM}) may
run over any complete set of basis spin states; the helicity eigenstate 
basis is a convenient choice. Consider the production of an electron and 
an (anti)neutrino with four-momenta and spins $p_1,s_1$ and $p_2, s_2$, 
respectively. 
For definiteness, we discuss the Dirac neutrino case, but our results will 
apply to the Majorana case as well. The initial and final leptonic states are 
\be
|i\rangle=|0\rangle\,,\qquad\qquad  
|f\rangle=b^\dag_e(p_1,s_1)d^\dag_\nu(p_2,s_2)|0\rangle\,.
\label{eq:if}
\ee
Substituting the expressions for $e(x)$ and $\nu(x)$ from
(\ref{eq:decompE}) and (\ref{eq:decompNuD}) into~(\ref{eq:leptCC}) and using
the standard anticommutation rules for fermionic creation and annihilation
operators, for the matrix element of the leptonic current one finds
\begin{align}
\langle f|j^\mu(x)|i\rangle 
=\bar{u}_e(p_1,s_1)\gamma^\mu(1-\gamma_5)v_\nu(p_2,s_2)
e^{i(p_1+p_2)x}\,.
\label{eq:mElem1}
\end{align}
The amplitude of the process is then obtained upon the convolution of the 
matrix element of the leptonic current 
(\ref{eq:mElem1}) with the matrix element of the hadronic current 
$\langle A_f|J_\mu|A_i\rangle$.

Now, an important point is that 

\noindent
{\sl {\color{blue}
the chirality projection operators 
can only act on spinors, such as $u$ and $v$, or on quantum field operators, 
which contain $u$ and $v$ spinors in their expansions, but not on the states. 
The latter, such as e.g.\ those given in eq.~(\ref{eq:if}), are vectors in a 
Hilbert space and do not have any spinorial structure. Chiral spinors only 
enter into the expressions for the amplitudes of neutrino production, but the 
amplitudes themselves are numbers, not spinors. The further evolution of the 
produced neutrino states is independent of the chiral structure of the 
interactions by which they were produced, though their production 
probabilities of course depend on it. 
}}

A comment on neutrino wave functions is in order. The quantity 
$(1-\gamma_5)v_\nu(p_2,s_2)e^{ip_2 x}=2v_{\nu R}(p_2,s_2)e^{ip_2 x}$\,%
\footnote{We follow the convention $P_{L,R}u=u_{L,R}$, $P_{L,R}v=v_{R,L}$.}  
in eq.~(\ref{eq:mElem1}) is actually twice the wave function of the produced 
antineutrino. However, its space-time dependence has no effect on the 
evolution of the antineutrino after its production. 
This comes about because in calculating the amplitude of the process 
one has to integrate over the production 4-coordinate $x$; 
the factor $e^{i p_2 x}$, together 
with the other similar factors in the matrix element of the process, 
just yields $\delta$-functions (or $\delta$-like functions) enforcing 
energy and momentum conservation \cite{Smirnov:2025wax}. 
The evolution 
of neutrinos following their production is thus described by the
evolution of their state vectors, not wave functions.

What determines  the evolution of neutrino states in vacuum? 
It is governed by the Hamiltonian of free neutrinos, 
which in QFT is given by the second-quantized expression 
\vspace*{1mm}
\be
H=\sum_{p,s}E_p[b^\dag(p,s)b(p,s)+d^\dag(p,s)d(p,s)\big)]\,,
\label{eq:H}
\vspace*{-2mm}
\ee
where the energy $E_p\equiv+\sqrt{{\vec{p}\,^2+m^2}}$ is non-negative. 
A one-particle neutrino state $|\nu(p_1,s_1)\rangle=b^\dag(p_1,s_1)
|0\rangle$ satisfies 
$H|\nu(p_1,s_1)\rangle=E_{p_1}|\nu(p_1,s_1)\rangle$, which gives   
\be 
e^{-i H t}|\nu(p_1,s_1)\rangle = e^{-i E_{p_1} t}|\nu(p_1,s_1)\rangle\,. 
\label{eq:evol}
\ee 
For Majorana neutrinos we have $H=\sum_{p,s}E_p a^\dag(p,s)a(p,s)$, 
and a similar consideration applies. 
Thus, the evolution of the produced neutrino states does not involve any 
phase factors corresponding to negative energies. Recall that the alleged 
existence of such factors was an important ingredient in the derivations 
of the probabilities of chirality oscillations. 

How can this be compared with the oscillations of neutrino flavour? In that 
case the oscillations are a consequence of the leptonic mixing, 
\be
\nu_a (x)=\sum_i U_{ai} \nu_i(x)\,,\qquad\quad
|\nu_a\rangle=\sum_i U_{ai}^* |\nu_i\rangle\,, 
\label{eq:mixing}
\ee
where the first equation describes the mixing of neutrino fields and the 
second one, the corresponding mixing of neutrino states. With time, the mass 
eigenstates $|\nu_i\rangle$ develop different phase factors $e^{-iE_i t}$, and 
the phase differences result in neutrino flavour oscillations. All 
the energies $E_i$ here are, of course, non-negative.  Note that the 
neutrino fields $\nu_a(x)$ and $\nu_i(x)$ are spinorial (and therefore can be 
chiral), whereas the states $|\nu_a\rangle$ and $|\nu_i\rangle$ are not. 
Let us stress once again, the oscillations arise due to nontrivial evolution 
of states, not of fields or wave functions; the similarity of the notation 
used for these objects is probably one of the main reasons of the confusion 
that led to the erroneous claims that neutrinos can undergo chirality 
oscillations.

\section{\label{sec:negative}Negative energies?} 

As was mentioned above, another crucial source of these claims were 
misconceptions about negative energies. 
This point deserves a more detailed discussion. 
The notion of negative energies comes from the 
relativistic single-particle equations of motion, which have solutions 
corresponding to both positive and negative energies; the latter are usually 
associated with antiparticles. The existence of negative-energy solutions 
leads to a number of perplexing consequences, such as 
e.g.\ {\em zitterbewegung} -- a fast oscillatory (``trembling") motion of 
particles (see e.g.~\cite{Itzykson:1980rh}). 

In contrast to this, there are no negative energies within the QFT framework, 
and no need for them. For example, in the single-particle approach 
{\em zitterbewegung} smears out the particle's coordinates, effectively 
delocalizing them to the region of the size of their Compton wavelength. 
In QFT, particles cannot be localized within regions of space 
smaller then their Compton wavelengths because an attempt to do this would 
lead to the production of pairs; no negative energies or ``trembling motion"  
are involved in this argument. 

That negative energies do not play any role (and actually do not exist) 
within QFT can be seen from the expansions of quantum field operators, 
such as those given by eqs.~(\ref{eq:decompE})-(\ref{eq:decompNuM}). 
The creation and annihilation operators enter in these expansions multiplied  
by the phase factors containing $+iEt$ and $-iEt$ in the exponents. This is 
simply because one of them creates a particle and therefore increases the 
energy of the system, whereas the other annihilates it and so decreases the 
energy. In both cases the added or subtracted energy is, of course, positive.  
There nothing more to it. 

The above argument has in general no relation with the existence of 
antiparticles. As can be seen from eq.~(\ref{eq:decompNuM}), it also applies 
to Majorana neutrinos, for which there is no distinction between particles and 
antiparticles. Likewise, it also applies to any other genuinely neutral 
particles, such as e.g.\ photons. 
The only thing that matters is whether a particle is created or annihilated, 
and not whether we call it particle or antiparticle. 

It is easy to see why the same arguments 
would not work in single-particle 
relativistic quantum mechanics. In that case the expressions of the type 
(\ref{eq:decompE})-(\ref{eq:decompNuM}) describe the most general solutions 
of the corresponding single-particle equations of motion, but the quantities 
like $b(p,s)$ and $d^*(p,s)$ (or $a(p,s)$ and $a^*(p,s)$) are simply complex 
numbers and not creation or annihilation operators. It is 
the field quantization that resolves the problem of negative energies. 

{\color{blue}{\sl 
Thus, negative energies are an artifact of insisting on using a 
single-particle description in particle physics. 
They are unphysical.}}

\bigskip 
To conclude this section, we quote from the two texts on QFT written by 
the prominent experts in the field. 

\vspace*{1.5mm}
{\em Notes from Sidney Coleman's Physics 253a \cite{Coleman:2011xi}:}

\noindent
``There are extensive discussions of the solutions of the Dirac equation 
in the literature. All you have to do is ignore all references to holes 
and negative solutions. Just because something was understood in a poor 
way 50 years ago, doesn't mean you have to learn it that way today.''

\vspace*{1.5mm}
{\em Anthony~Zee, Quantum Field Theory in a Nutshell (2nd edition) 
\cite{Zee:2010qce}:}

\noindent
``In closing this chapter let me ask you some rhetorical questions. Did I 
speak of an electron going backward in time? Did I mumble something 
about a sea of negative energy electrons? This metaphorical language, 
when used by brilliant minds, the likes of Dirac and Feynman, was 
evocative and inspirational, but unfortunately confused generations of 
physics students and physicists. The presentation given here is in the 
modern spirit, which seeks to avoid these potentially confusing 
metaphors.''

\section{\label{sec:chiral}Chiral states?} 
We argued in sec.~\ref{sec:QFT} that the notion of chirality applies only to  
spinorial objects, such as the solutions of the Dirac equation or fermionic 
quantum field operators, but not to the states. How can then one understand 
the frequently found in the literature statements that neutrinos are created 
(and absorbed) via the charged-current weak interactions  in the states of 
left-handed chirality and their antiparticles in the states  
of right-handed chirality?

This is actually a jargon, which means that,  
due to the $V-A$ structure of the leptonic current, 
the amplitudes of production and detection of neutrinos and 
antineutrinos\,%
\footnote{Here and in the discussion below in the Majorana neutrino case by 
``antineutrinos" we mean neutrinos produced e.g. in $\beta^-$ decays or 
detected via inverse 
$\beta^-$ decays, i.e.\ participating in the processes in which in the 
Dirac case only antineutrinos could take part. Examples are neutrinos 
produced in nuclear reactors and detected via inverse $\beta$-decay on 
protons.}
in the states of positive and negative helicities 
$(h=\pm 1$) are proportional to the factors%
\be
f_{\nu h}=1-h\frac{p}{E+m}\,, \qquad\qquad
f_{\bar{\nu} h}=1+h\frac{p}{E+m}\,,
\label{eq:f}
\ee
where $p\equiv|\vec{p}|$. In particular, for ultra-relativistic neutrinos 
the produced or detected states should predominantly have negative 
helicity, whereas the amplitudes of production or detection of 
positive-helicity neutrinos are strongly suppressed. For antineutrinos the 
situation is opposite. 
At the same time, in the nonrelativistic case the factors $f_{\nu h}$ and 
$f_{\bar{\nu} h}$ are practically the same for neutrinos and antineutrinos 
of either helicity. 

The terms ``left-handed" and ``right-handed", when used to describe 
neutrino states, 
are therefore simply shorthands serving to avoid lengthy explanations. 

As any jargon, however, this one also has its limitations. We pointed 
out above that the production and detection amplitudes for neutrinos or 
antineutrinos of helicity $h$ are proportional to the factors $f_{\nu h}$ and 
$f_{\bar{\nu} h}$. This does not in general mean, however, that {\sl all} 
the dependence of the amplitudes on $h$ is fully 
contained in these factors. The amplitudes of the weak interaction processes 
are obtained upon the convolution of the matrix elements of the leptonic 
current with the matrix elements of the current of the other participating 
particles. In particular, in the case of nuclear $\beta$-decay these are the 
hadronic matrix elements $\langle A_f|J_\mu|A_i\rangle$. 
This can leave an imprint on the dependence of the amplitudes of the process 
on neutrino helicity; this dependence can also be influenced by the 
spin states and angular distributions of the accompanying charged leptons. 
The proportionality to the factors $f_{\nu h}$ or $f_{\bar{\nu} h}$ 
will still remain, but the overall dependence on $h$ can be modified 
significantly. For the special case of neutron $\beta$-decay this issue has 
been studied in detail in \cite{Smirnov:2025wax}. 

Although up to now we have only been discussing neutrino chirality,  
the same analysis should also apply to charged leptons participating 
in the same charged-current weak processes in which neutrinos are produced 
or detected. There is, however, at least one case in which such considerations 
would not be fully adequate: 2-body weak decays 
of charged pseudoscalar mesons, such as e.g.\ $\pi^+\to \mu^++\nu_\mu$. 
In the rest frame of the parent pion the produced muon and neutrino  
fly away back to back, and due to angular momentum conservation 
they must have the same helicities. If one neglects neutrino mass, the 
produced neutrino will be in the pure negative helicity state, and so must 
be $\mu^+$. This is actually opposite to what one would expect for a 
relativistic antilepton produced by a $V-A$ interaction. 
One can define for charged leptons the factors similar to those given for 
neutrinos in eq.~(\ref{eq:f}); the amplitude of the process with the 
production of $\mu^+$ should then be proportional to 
$f_{\bar{\mu}h_\mu}=[1+h_\mu p_\mu/(E_\mu+m_\mu)]$ 
(in the obvious notation).  
However, this expression does not describe the relative magnitudes of the 
amplitudes of the process with 
$h_\mu=+1$ and $-1$; as noted above, the value $h_\mu=+1$ is strictly 
forbidden by angular momentum conservation. Still, for $h_\mu=-1$ the factor 
$f_{\bar{\mu} h_\mu}$ correctly reproduces the fact that the amplitude 
of the $\pi^+\to \mu^++\nu_\mu$ decay is proportional to the muon mass. 
A recent analysis of such decays in the general case when the mass of 
neutrinos is not neglected can be found in \cite{Smirnov:2025wax}. 

To summarize, the terms ``left-handed neutrinos" and ``right-handed 
antineutrinos" apply to the spinors ($u_L$ and $v_R$) associated with 
neutrino and antineutrino wave functions and entering in the expansions of 
neutrino quantum field operators as well as in the expressions for the 
amplitudes of neutrino production and detection. By contrast, the notion of 
chirality does not apply to neutrino states, which, ignoring the flavour 
issues, are characterized only by neutrino momenta and spin states. 
Oscillations of neutrino chirality in vacuum, which would require production 
and detection of ``chiral neutrino states'', do not occur.

\vspace*{1.0mm}
{\em Acknowledgements.} The author gratefully acknowledges very useful 
discussions with Georg Raffelt and Alexei Smirnov.

\end{document}